\documentclass[a4paper]{spie}  %>>> use this instead for A4 paper
\usepackage{amsmath,amsfonts,amssymb}
\usepackage{graphicx}
\usepackage[colorlinks=true, allcolors=blue]{hyperref}

\title{The calibration procedure of the LINC-NIRVANA ground
and high layer WFS}

\author[a,d]{Carmelo Arcidiacono}
\author[b]{Kalyan Kumar Radhakrishnan Santhakumari}
\author[c,d]{Valentina Viotto}
\author[c,d]{Maria Bergomi}
\author[b]{Florian Briegel}
\author[b]{Thomas Bertram}
\author[c,d]{Luca Marafatto}
\author[b]{Tom Herbst}
\author[c,d]{Jacopo Farinato}
\author[c,d]{Roberto Ragazzoni}
\author[b]{Ralph Hofferbert}
\author[b]{Martin K\"urster}
\author[b]{Frank Kittman}
\author[b]{J\"urgen Berwein}
\author[b]{Harald Baumeister}

\affil[a]{INAF -- Osservatorio Astrofisico e scienza dello Spazio di Bologna, Via P. Gobetti 93/3, 40129 Bologna, Italy}
\affil[b]{MPIA -- Max-Plank-Institut f\"ur Astronomie, K\"onigstuhl 17, 69117 Heidelberg, Germany}
\affil[c]{INAF -- Osservatorio Astronomico di Padova, Vicolo dell'Osservatorio 5, 35127 Padova, Italy}
\affil[d]{ADONI -- Laboratorio Nazionale di Ottica Adattiva, Italy}
\authorinfo{Further author information: (Send correspondence to Carmelo Arcidiacono)\\C.A.: E-mail: carmelo.arcidiacono@inaf.it, Telephone:  +39 051 6357 316}

\begin{document} 
\maketitle

\begin{abstract}
LINC--NIRVANA (LN) is an MCAO module currently mounted on the Rear Bent Gregorian focus of the Large Binocular Telescope (LBT). It mounts a camera originally designed to realize the interferometric imaging focal station of the telescopes. LN follows the LBT binocular strategy having two twin channels: a double Layer Oriented Multi-Conjugate
Adaptive Optics system assisting the two arms,
supplies high order wave-front correction.
In order to counterbalance the field rotation, a mechanical derotation is applied for
the two ground wave-front sensors, and an optical (K-mirror) one for the two high
layers sensors, fixing the positions of the focal planes with respect to the
pyramids aboard the wavefront sensors.
The derotation introduces a pupil images rotation on the wavefront
sensors, changing the projection of the deformable mirrors on the sensor
consequently.

\end{abstract}


\section{INTRODUCTION}
\label{sec:intro}  % \label{} allows reference to this section

LINC--NIRVANA\cite{2003SPIE.4839..536R} (LN) has been designed to realize the interferometric imaging focal station of the Large Binocular Telescope\cite{2010ApOpt..49..115H} (LBT). It is composed by a Fizeau interferometry ready camera and a double Layer--Oriented\cite{LO1,LO2,arcidiacono07} (LO) multi-conjugate adaptive optics\cite{beckers88,beckers89a} (MCAO) system. The MCAO module assists the two arms of the interferometer\cite{2010SPIE.7734E..3MV}, supplying a high order wave-front correction.
In particular, in LN we implement the Multiple Field of View\cite{mfov,2008SPIE.7015E.149F} (MFOV) a version of the LO, driving independently the loops for the ground and the high layers correction respectively.
According to the LO scheme, we realize two independent loops using two separate Wavefront sensors (WFSs), collecting the light of up to 12 reference stars for the Ground WFS (GWS) and up to 8 for the high (HWS). The Field of View  (Fov) of the two WFSs are different: a 1~{arc min} to 3~{arc min} FoV radii annulus for the GWS and a full 1~{arc min} radius FoV for the HWS.
In both WFS the sensing implements the optical co-addition of several Pyramid\cite{pyramid} sensors: the light of all the natural reference stars within the FoV is split by as many pyramids into four beams. Each beam generates a pupil image through a common objective: the pupils corresponding to different references finally overlap on a WFS mimicking the geometrical overlap of the footprints of the stars beam in the atmosphere at the desired conjugation altitude: 0~km and 7.1~km for the GWS and HWS respectively. The deformable mirrors complete the conceptual design of the adaptive optics system, being the Adaptive Secondary\cite{riccardi2003} (AdSec) coupled to the GWS\cite{2012SPIE.8447E..0VC} and the 349-actuator piezo-stack deformable mirror from Xinetics mounted on the LN's bench for the HWS\cite{2011OExpr..1916087Z}.

\begin{figure} [ht]
   \begin{center}
   \begin{tabular}{c}
   \includegraphics[height=6.2cm]{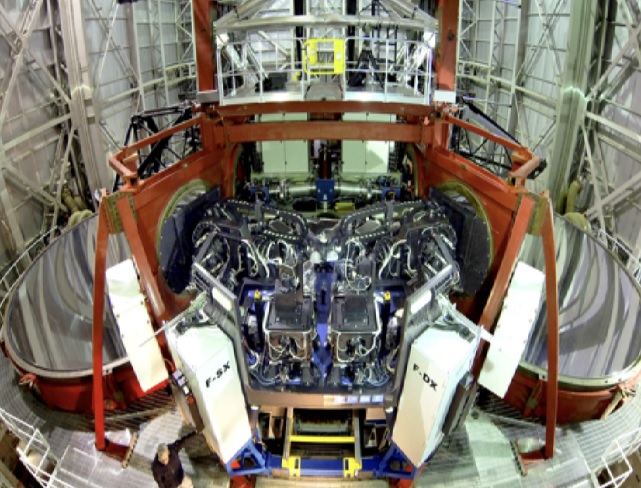}
   \includegraphics[height=6.2cm]{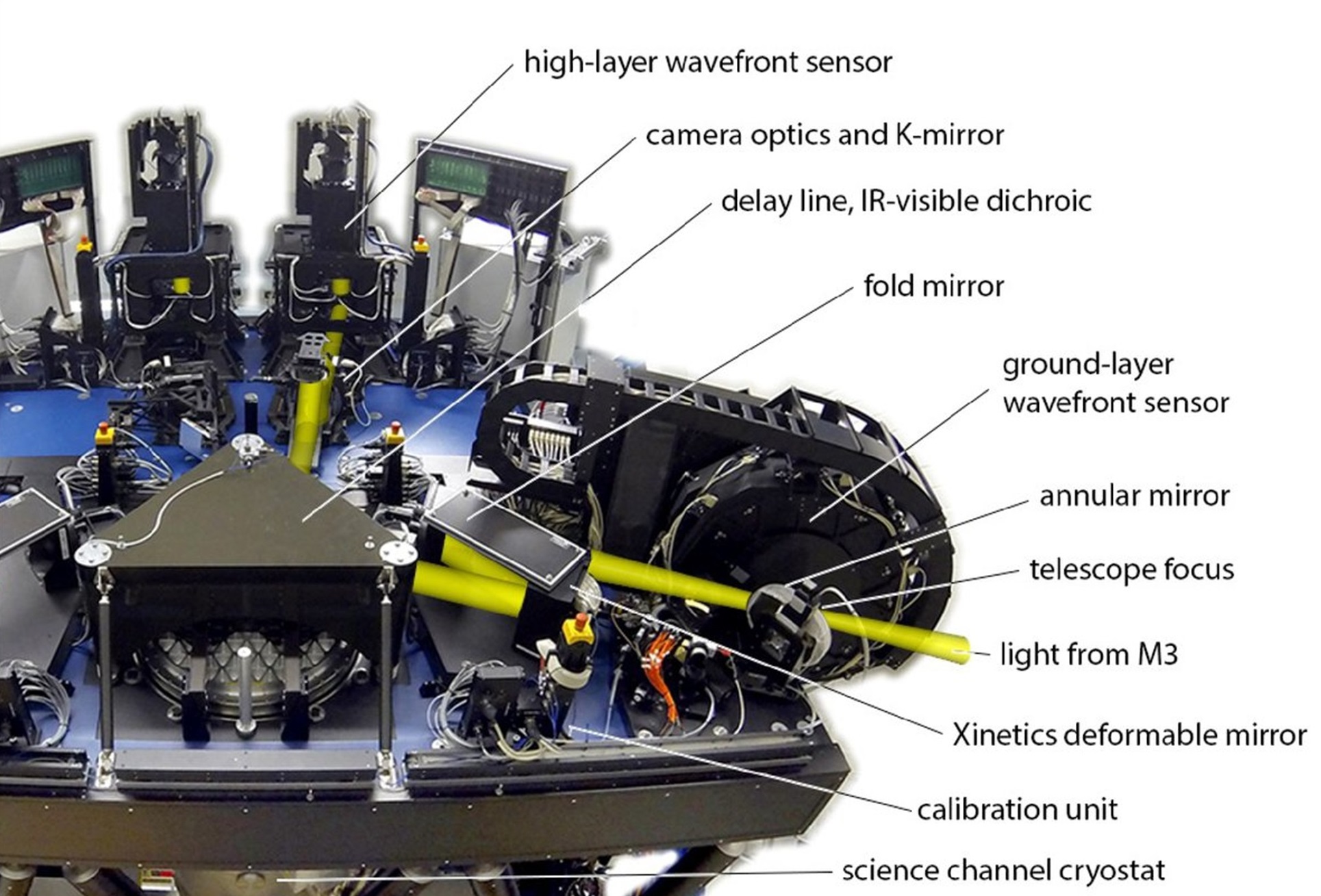}
	\end{tabular}
	\end{center}
   \caption[example] 
   { \label{fig:figure0} 
The picture on the left shows the LN instrument mounted in the central platform of the LBT and, on the right, a CAD sketch of the instrument hosting on the top part the AO WFS: the HWSs are the two towers on the front and the GWSs are the elements at the edge of the bench on the left and right side. Finally, the infrared camera stands below the bench.}
   \end{figure}

LN sits on the rear, shared, bent Gregorian foci of the LBT: the LBT's M3 folding mirror forwards the light on a Field selector mirror on board the LN separating the two MFoV path to the GWS and HWS, see Figure~\ref{fig:figure0}. This system is implemented symmetrically on the two side of the LBT.
   
We mounted the two GWS on a rotation bearing, in order to counterbalance the field rotation,  and the placed a K mirror for the optical derotation of the HWS sensors. In this way, LN fixes the positions of the focal planes with respect to the pyramids aboard each wavefront sensor. However, the derotation introduces the rotation of the pupil image on both wavefront sensors: the projection of the deformable mirrors on the sensor consequently changes. The proper rotation correction of the control matrix will be applied in real-time through the upload of a new numerically computed matrix. 
Here we present the calibration strategy for the calibration of the GWS and HWS DM-WFS interaction matrices.

\section{Status}
As of June 2018, the commissioning of LN is ongoing at the Large
Binocular Telescope (LBT): the left arm has been fully operated in MCAO mode producing the first science images\cite{TomSpie2018}. 
We already discussed in \cite[Arcidiacono et al., 2010]{2010SPIE.7736E..4JA} of the error introduced by a not proper matching of the rotation angles of the WFS grid w.r.t. the actuator pattern on the DM and in \cite[Bertram et al., 2014]{2014SPIE.9148E..5MB} about the generation of the reconstructor matrix. Here we focus on the calibration procedure we follow to finally build the control matrix (reconstructor), that we need to upload it in the RTC to avoid the injection of WF error in the loops because of rotation of the DM-WFS mapping. The calibration procedure has been completed for the GWS and HWS on the left (SX) arm of the telescope. The calibration of the right side is still pending since a refurbishment of the AdSec secondary is expected in summer 2018.

\section{WFS and DMs}
   \begin{figure} [ht]
   \begin{center}
   \begin{tabular}{c} 
   \includegraphics[height=6cm]{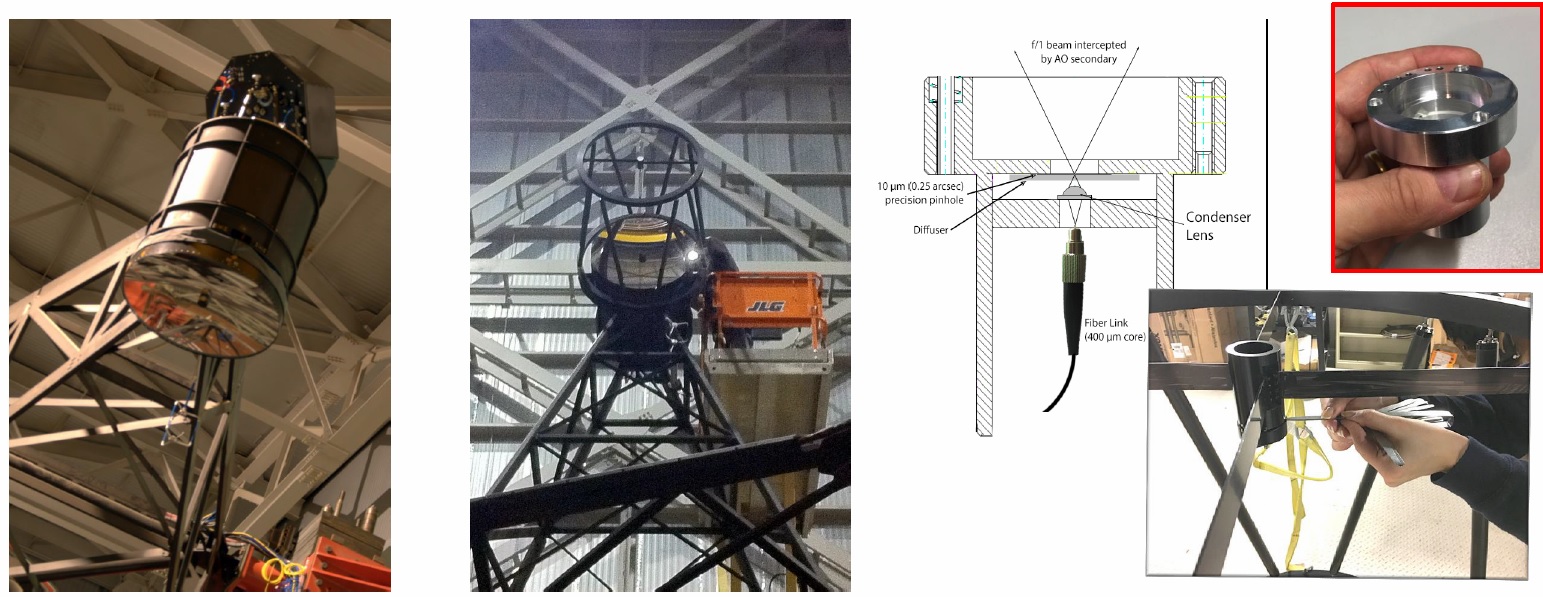}
	\end{tabular}
	\end{center}
   \caption[example] 
   { \label{fig:figure1} 
From left to right: the Adaptive Secondary (AdSec) of the LBT; the Retro-Reflector mounted on the AdSec; the light source: a fiber mounted on the focus of the AdSec.}
   \end{figure} 

The calibration procedure of the HWS and GWS are conceptually identical and, for the sake of space, we limit the calibration description mostly to the GWS.

The Ground GWS is a multi-pyramid (12) WFS coupled to the Adaptive Secondary (AdSec) of the LBT. In the HWS the pyramids are 8 and the sensor is optically coupled to a 349 actuator DM from Xinetics. 

Following the Layer Oriented scheme, the light of all the reference stars, once split by the corresponding refractive pyramid, is imaged on a single detector (CCD50 and CCD39 respectively for GWS and HWS) properly focused on the conjugated DM (the AdSec is the pupil of the telescope the Xinetics is conjugated to 7.1km). 
The CCDs are mounted on a 3 linear stages configuration that allows XYZ precision shifts.

\begin{figure} [ht]
   \begin{center}
   \begin{tabular}{c} 
   \includegraphics[height=5cm]{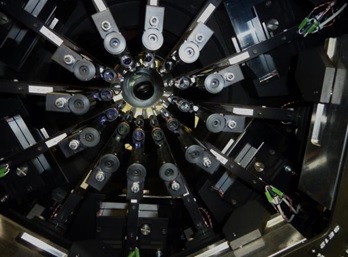}
   \includegraphics[height=5cm]{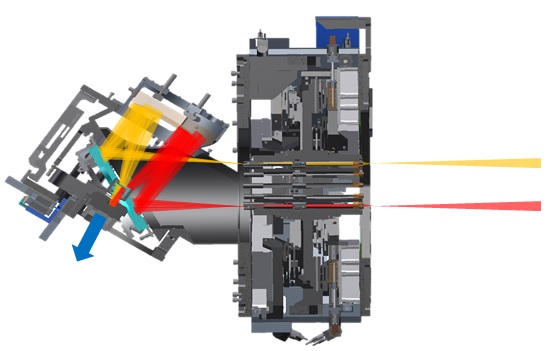}
	\end{tabular}
	\end{center}
   \caption[example] 
   { \label{fig:figure2} 
A scheme of the GWS: on the left the 12 pyramids, each one being hosted on board of a star enlarger, on the right a cross section of the GWS sensor. The optical path of two reference stars is enhanced. The light entering from the right side is forwarded to the pyramids placed on an intermediate focal plane.}
   \end{figure}

Each pyramid is mounted at the focus of an optical relay (a.k.a. star enlarger) moving the focal ratio from 15 to 187.5 (F/20 to F/300 for HWS). In this way, the GWS projects a pupil image diameter of 48 pixels on the 128$\times$128 pixels CCD50, see Figure~\ref{fig:figure2} and Figure~\ref{fig:4pupils}. 
For the GWS LN is not autonomous since we need to illuminate the AdSec to perform the calibration. However we have the chance to perform daytime calibrations by installing an optical fiber source placed mechanically on the focus of the AdSec, see Figure~\ref{fig:figure1}, avoiding the use of  night time.

The fiber is mounted trough a mounting, a.k.a. retro-reflector unit, originally designed to perform the calibration of the LBTI and FLAO WFS calibrations\cite{2010ApOpt..49G.174E}. This unit was reversibly modified to host our calibration light source. Actually, after our modification the name retro-reflector may be misleading since we placed a fiber holder instead of the original 
optical system composed by a F/1 parabola and a flat mirror.

For the GWS calibrations, we used the fiber source placed mechanically on the focus of the AdSec. According to the LO scheme in the GWS all the pyramids see the AdSec in the same way and are projected on the detector on the WFS in order to superimpose:
we cross-checked that we measure identical WFS-to-Deformable Mirror Interaction Matrix for all the pyramids, as we expected.

\begin{figure} [ht]
   \begin{center}
   \begin{tabular}{c} 
   \includegraphics[height=5cm]{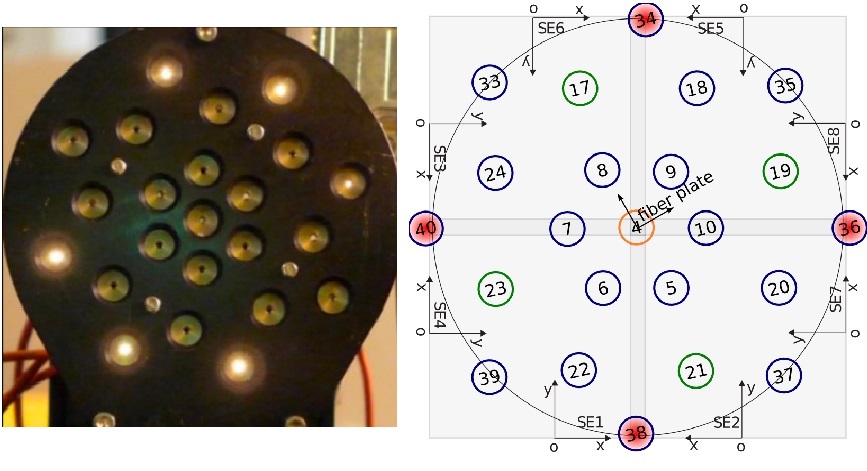}
	\end{tabular}
	\end{center}
   \caption[example] 
   { \label{fig:figureFP} 
The fiber plate composing part of the calibration unit of LN. We placed a set of 8 fiber in the most external slots, at a diameter of 1.85 FoV, for the HWS calibration.}
   \end{figure}

\begin{figure} [ht]
   \begin{center}
   \begin{tabular}{c} 
   \includegraphics[height=5cm]{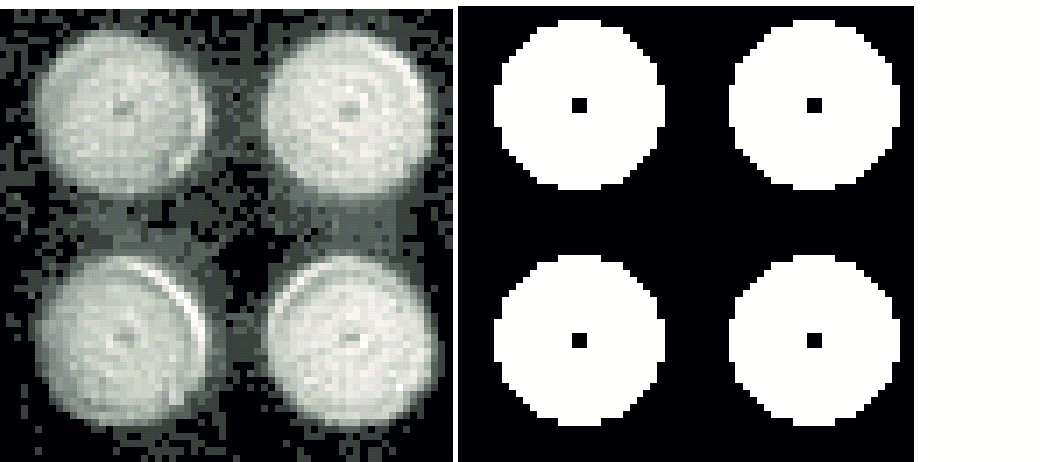}
   \includegraphics[height=5cm]{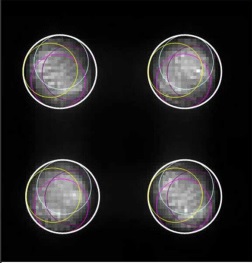}
	\end{tabular}
	\end{center}
   \caption[example] 
   { \label{fig:4pupils} 
On the left, a frame from the the CCD50 on the GWS WFS looking on a single fiber with AdSec applying the flat shape. On the central panel: the pupil mask as encoded in the Real Time Computer (RTC) software. Here the system used a 2$\times$2 binning on chip, the same used for on sky operations in the case of bright reference stars. On the right the CCD39 is focused in order to make the super imposition of the pupil covering the metapupil (as on the Xinetics DM).}
   \end{figure}

The HWS is similar to the GWS, the main difference being the conjugation altitude that in the HWS case is 7.1 km, optically conjugated to the Xinetics DM. In the HWS the footprints of the stars overlap in a way to mimic the geometry they got on a layer at 7.1~km altitude.
For the HWS  we used a set of optical fibers inserted in a intermediate focal plane. The light of 8 fibers was used to project the 2~{arc min} FoV metapupil to illuminate the DM, seen simultaneously by as many pyramids. Actually the illuminated FoV was undersized since we placed the fiber on a diameter corresponding to 1.85~{arc min}(see Figure \ref{fig:figureFP}).

In both GWS and HWS, 
each system pyramid+star enlarger is mounted on a couple of linear stages that allow the patrolling of the WFS FoV. The optical system of each pyramid unit is required to produce images of the pupil with an error on the diameter less than 1/10 of sub-aperture.  Let us remind that a pixel on the detector corresponds to a sub-aperture on the LO-pyramid WFS.
Given the small wavefront error\cite{2012SPIE.8447E..6FM} (WFE) introduced by the optical relay in front of the pyramid we may measure, in the GWS, almost identical WFS-Deformable Mirror Interaction Matrix for all the pyramids.

\section{Instrumental Setup}
Being the GWS interaction Matrix (IM) of the different pyramids interchangeable, we measured it using only one. 
We illuminated the pyramid projecting an F/1 beam on the AdSec shell, placing a fiber assisted optical system on one focus of the concave elliptical mirror (the AdSec). In order to simulate the effect of a partially corrected PSF (the typical reference is not diffraction limited since WFS work in the visible wavebands) we inserted a couple diffuser/pin-hole that generates a disk uniformly illuminated of 0”.5~{arc sec} on the pyramid pin. Special attention was taken to generate a uniform illumination of the mirror shell in order to not inject errors in the IM measurement process, since we want to simulate the light from a real star, that will produce a uniformly illuminated pupil image. The CCD binning for the foreseen night-time operations is set to two, making 64$\times$64 the size of the images on which we measure the four pupil differential illumination. Using the binning 2 of the detector the slopes are measured on 24px diameter pupils.

Both GWS and HWS CCDs are mounted on a couple of linear stages\cite{2004SPIE.5490.1286S} positioned one on top of the other in a 90 degrees configuration. The detector linear stages are remotely controlled, allowing an active centering of the detector with respect to the metapupil image.

\section{Measurements Preparation}
In order to start with the measurements of the IM, we needed to properly center the pupil images with respect to the mask used by the RTC for selecting the valid sub-aperture from the CCD images.
Since we need to perform rotation of the IM matrix to take into account for optical rotation of AdSec and Xinetics with respect to the detector, we needed  to identify the center of rotation of the mirror images on the CCD. We had to move that measured center as close as possible to the center of the circular pupil we draw for the RTC mask.
This was an iterative process in which we use as reference the signal of the first 100 modes of the Karhunen-Loeve (KL) modal base used to control the mirror for different position of the detector linear stages. In this way, and independently on the optical illumination of the pupil, we succeeded to perform a centering. 
In a later stage, as part of the acquisition procedure, we will need to center the pupil image with respect to the four pupil image geometry obtained by looking into the mirror modes shape in this calibration phase.
We perform this later centering with a precision of about 1/10 of sub-aperture. 

From the operational point of view, we built a python script that takes care of the measurements by preparing the KL modes (the modes to command, M2C, matrix), defining the push-pull history file to be uploaded to the AdSec Basic Computation Unit (BCU) on board of the AdSec\cite{2003SPIE.4839..772B} or to the other BCU in the case of HWS. The script interacts with the AdSec/Xinetics in order to finally download the measured slopes history file corresponding to the applied shapes.
The script takes care of the slopes measurement extraction and of the generation of a IM sample corresponding to about 12-16 repetitions of the push-pull sequence for each mode. Finally the script averages the IMs generated in this way and finally compute the average one.

\section{IM Calibrations}
To optimize the result of the calibration process it is required that pyramids work within the linear regime. Lower the WF aberration experienced by the pyramids the higher would be the sensitivity. For the IM measurements we followed an iterative process to flat the WF in front of the sensor. For the initial measurements we needed to save IM with 2, 5, 10, 20, 50, 70, 90 and 100 modes and as many reconstructors in order to build a well performing optical flat of the deformable mirror. This fine tuning process is needed at the very beginning of the calibration procedure only.

Once this best shape was achieved, we save this shape as reference flat of the DM. Later, as part of the calibration measurement, we perform a minor revision of the flat, adjusting the AdSec command vector by closing an AO loop with 50 modes and saving the average mirror shape that we will use for the following steps as new DM-flat.

For each calibration run we save a set of IMs registering 2 (tip and tilt), 50 and 100modes. Actually, because of the unavoidable vibrations, the measurement of the tip-tilt were more noisy than other modes. And a special attention to these two modes was necessary, averaging a much larger of frame according to the actual level of noise. 

\section{Different Angles}
As said in the introduction, LINC-NIRVANA WFSs need to follow sky rotation to keep the pyramids properly centered on the reference stars focal plane images. This generates the rotation of the DM actuator pattern on the detector. Actually, the detectors see the rotation of both the actuators map and the slopes directions.

   \begin{figure} [ht]
   \begin{center}
   \begin{tabular}{c} 
   \includegraphics[height=18cm]{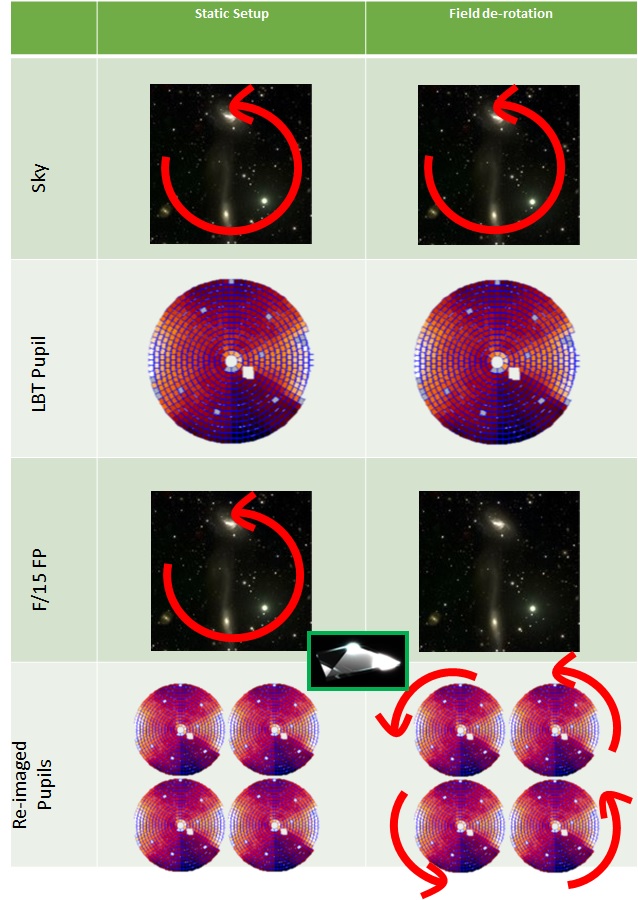}
	\end{tabular}
	\end{center}
   \caption[example] 
   { \label{fig:fieldrotation} 
On the left column the effect of the sky rotation in static condition leaving  the bearing off-line. On the right the bearing follows sky rotation but in this way produces the rotation of the pupil as seen from the WFSs.}
   \end{figure} 

Since the Influence Function of the single actuator is not properly sampled using 24 sub-aperture across the pupil (we have 28 actuators along the diameter of the AdSec) we suffer of under-sampling. Because of this, some couple of actuators fall inside one sub-aperture making the WFS insensitive to possible slopes in between the two. The numbers are more equilibrated in the HWS case since we have 21 actuators across the diameter, just slightly smaller of the number of valid sub-apertures on the detector (about 26 at the finest sampling).

One of the challenges is that the sensitivity of the WFS to the actuator pattern changes slightly with the rotation.

To recover a better resolution and having in mind to compute the average sensitivity on all the actuators, we need to perform more measurements for different positions of the actuator pattern w.r.t. the grid of sub-apertures on the detector. Because of geometrical reasons (the GWS has annular patterns, while Xinetics DM use a square grid geometry with sampling slightly smaller than the sub-aperture projected one) and since the challenge we want to deal is produced by the pupil rotation, we choose to rotate instead of, for example, moving in X and Y the detector in the WFS. We measured 5 different interaction matrices for as many rotation angles. In this way we measured the effect of each actuator: we performed rotation steps not equally angularly spaced and covering 10degree of rotation avoiding effects related to the periodicity of the actuators pattern.

Because of the noise induced by vibrations we needed to perform separately the tip-tilt measurements averaging, typically, 50 times more push-pulls couples than what we did for the other modes.
For each angle we saved a 50modes IM and we used this to adjust the flat of the AdSec to finally measures the 100modes IM.
Once we have the 5 interaction matrices, we counter-rotates them – rotating the map of the X and Y slopes and then rotating the slopes X and Y vector of the corresponding angle needed to match the angle “zero” for the rotation. We then average them and this average-IM is back-rotated to the various angles needed. 

Our on-sky strategy foresees to update the control matrix (100modes) for every 1degree of rotation in order to do not affect the performance. The angle “zero” IM is then rotated for the full range made available by the GWS bearing and HWS K-mirror.

In the case of the GWS the computation of the control matrices is performed off-line and the full set of possible angles is stored on the LN servers. During operation the RTC system is started having loaded the first two reconstructor expected. The matrix vector multiplication of the control matrix by the slope vector is performed on the BCU\cite{2011aoel.confP..44B} on board of the AdSec.

\section{Difference GWS-HWS}
The calibration procedure for the HWS is very similar to the GWS one, with a few exception. For the HWS have installed a calibration unit that can be inserted in the optical path inserting a folding mirror. On this unit we have the possibility to install the 8 fibers and add an additional IR fiber to illuminate the IR channel.  Let us remark that using one single fiber in the case of the HWS is not producing the desired measurements since the metapupil is only partially filled. On the other side, we have the possibility to turn on all the fibers simultaneously making the IM measurements possible.
In the case of the HWS the rotation of the pupil is not directly affecting the measurements since the controlled DM is the internal one. However it is clocked with the pupil making all the description made for the GWS valid also for the HWS.
\section{Conclusions}
We described the calibration strategy for the GWS (and HWS) of the LINC-NIRVANA. The complication of the rotation of the DM as seen by the WFS was solved by averaging different interaction matrix recorded at different angles. The interaction matrix computed in this way were successfully used both in the laboratory and on sky\cite{2014SPIE.9148E..2YB} and both for the GWS and HWS.
As of June 2018 the LN is being commissioned\cite{2016SPIE.9909E..2UH,TomSpie2018} at the LBT observatory.

\acknowledgments % equivalent to \section*{ACKNOWLEDGMENTS}       
The authors wish to thank the LBTO and in particular the LBT crew that supported all our activities at the telescope.

% References
\bibliography{report} % bibliography data in report.bib

\begin{thebibliography}{10}

\bibitem{2003SPIE.4839..536R}
{Ragazzoni}, R., {Herbst}, T.~M., {Gaessler}, W., {Andersen}, D.,
  {Arcidiacono}, C., {Baruffolo}, A., {Baumeister}, H., {Bizenberger}, P.,
  {Diolaiti}, E., {Esposito}, S., {Farinato}, J., {Rix}, H.~W., {Rohloff},
  R.-R., {Riccardi}, A., {Salinari}, P., {Soci}, R., {Vernet-Viard}, E., and
  {Xu}, W., ``{A visible MCAO channel for NIRVANA at the LBT},'' in [{\em
  Adaptive Optical System Technologies II}{\nolinebreak\hspace{0.1em}]},
  {Wizinowich}, P.~L. and {Bonaccini}, D., eds., {\em \procspie} {\bf 4839},
  536--543 (2003).

\bibitem{2010ApOpt..49..115H}
{Hill}, J.~M., ``{The Large Binocular Telescope},'' {\em Applied Optics}~{\bf
  49},  115--122 (2010).

\bibitem{LO1}
{Ragazzoni}, R., ``{Adaptive optics for giant telescopes: NGS vs. LGS},'' in
  [{\em Proceedings of the Backaskog workshop on extremely large
  telescopes}{\nolinebreak\hspace{0.1em}]},  {Andersen}, T., {Ardeberg}, A.,
  and {Gilmozzi}, R., eds.,  175--180 (2000).

\bibitem{LO2}
{Ragazzoni}, R., {Farinato}, J., and {Marchetti}, E., ``{Adaptive optics for
  100-m-class telescopes: new challenges require new solutions},'' in [{\em
  Adaptive Optical Systems Technology}{\nolinebreak\hspace{0.1em}]},
  {Wizinowich}, P.~L., ed., {\em Proc. SPIE} {\bf 4007},  1076--1087 (2000).

\bibitem{arcidiacono07}
{Arcidiacono}, C., {Lombini}, M., {Farinato}, J., and {Ragazzoni}, R.,
  ``{Toward the first light of the Layer Oriented Wavefront Sensor for MAD.},''
  {\em Memorie della Societa Astronomica Italiana}~{\bf 78},  708--711 (2007).

\bibitem{beckers88}
{Beckers}, J.~M., ``{Increasing the size of the isoplanatic patch with
  multiconjugate adaptive optics.},'' in [{\em ESO Conference on Very Large
  Telescopes and their Instrumentation}{\nolinebreak\hspace{0.1em}]},   {\bf
  2},  693--703 (1988).

\bibitem{beckers89a}
{Beckers}, J.~M., ``{Detailed compensation of atmospheric seeing using
  multiconjugate adaptive optics.},'' in [{\em Active Telescope
  Systems}{\nolinebreak\hspace{0.1em}]},  {\em Proc. SPIE} {\bf 1114},
  215--217 (1989).

\bibitem{2010SPIE.7734E..3MV}
{Viotto}, V., {Ragazzoni}, R., {Arcidiacono}, C., {Bergomi}, M., {Brunelli},
  A., {Dima}, M., {Farinato}, J., {Gentile}, G., {Magrin}, D., {Cosentino}, G.,
  {Diolaiti}, E., {Foppiani}, I., {Lombini}, M., {Schreiber}, L., {Bertram},
  T., {Bizenberger}, P., {De Bonis}, F., {G{\"a}ssler}, W., {Herbst}, T.,
  {Kuerster}, M., {Meschke}, D., {Mohr}, L., and {Rohloff}, R.-R., ``{A very
  wide field wavefront sensor for a very narrow field interferometer},'' in
  [{\em Optical and Infrared Interferometry II}{\nolinebreak\hspace{0.1em}]},
  {\em \procspie} {\bf 7734},  77343M (2010).

\bibitem{mfov}
{Ragazzoni}, R., {Diolaiti}, E., {Farinato}, J., {Fedrigo}, E., {Marchetti},
  E., {Tordi}, M., and {Kirkman}, D., ``{Multiple field of view layer-oriented
  adaptive optics. Nearly whole sky coverage on 8 m class telescopes and
  beyond},'' {\em \aap}~{\bf 396},  731--744 (2002).

\bibitem{2008SPIE.7015E.149F}
{Farinato}, J., {Ragazzoni}, R., {Arcidiacono}, C., {Brunelli}, A., {Dima}, M.,
  {Gentile}, G., {Viotto}, V., {Diolaiti}, E., {Foppiani}, I., {Lombini}, M.,
  {Schreiber}, L., {Bizenberger}, P., {De Bonis}, F., {Egner}, S.,
  {G{\"a}ssler}, W., {Herbst}, T., {K{\"u}rster}, M., {Mohr}, L., and
  {Rohloff}, R., ``{The Multiple Field of View Layer Oriented wavefront sensing
  system of LINC-NIRVANA: two arcminutes of corrected field using solely
  Natural Guide Stars},'' in [{\em Adaptive Optics
  Systems}{\nolinebreak\hspace{0.1em}]},  {\em Proc. SPIE} {\bf 7015} (2008).

\bibitem{pyramid}
{Ragazzoni}, R., ``{Pupil plane wavefront sensing with an oscillating prism},''
  {\em Journal of Modern Optics}~{\bf 43},  289--293 (1996).

\bibitem{riccardi2003}
Riccardi, A., Brusa, G., Salinari, P., Gallieni, D., Biasi, R., Andrighettoni,
  M., and Martin, H.~M., ``Adaptive secondary mirrors for the {L}arge
  {B}inocular {T}elescope,'' {\em Proceedings of SPIE}~{\bf 4839},  721--732
  (2003).

\bibitem{2012SPIE.8447E..0VC}
{Conrad}, A.~R., {Arcidiacono}, C., {Baumeister}, H., {Bergomi}, M., {Bertram},
  T., {Berwein}, J., {Biddick}, C., {Bizenberger}, P., {Brangier}, M.,
  {Briegel}, F., {Brunelli}, A., {Brynnel}, J., {Busoni}, L., {Cushing}, N.,
  {De Bonis}, F., {De La Pena}, M., {Esposito}, S., {Farinato}, J., {Fini}, L.,
  {Green}, R.~F., {Herbst}, T., {Hofferbert}, R., {Kittmann}, F., {Kuerster},
  M., {Laun}, W., {Meschke}, D., {Mohr}, L., {Pavlov}, A., {Pott}, J.-U.,
  {Puglisi}, A., {Ragazzoni}, R., {Rakich}, A., {Rohloff}, R.-R., {Trowitzsch},
  J., {Viotto}, V., and {Zhang}, X., ``{LINC-NIRVANA Pathfinder: testing the
  next generation of wave front sensors at LBT},'' in [{\em Adaptive Optics
  Systems III}{\nolinebreak\hspace{0.1em}]},  {\em \procspie} {\bf 8447},
  84470V (2012).

\bibitem{2011OExpr..1916087Z}
{Zhang}, X., {Gaessler}, W., {Conrad}, A.~R., {Bertram}, T., {Arcidiacono}, C.,
  {Herbst}, T.~M., {Kuerster}, M., {Bizenberger}, P., {Meschke}, D., {Rix},
  H.-W., {Rao}, C., {Mohr}, L., {Briegel}, F., {Kittmann}, F., {Berwein}, J.,
  {Trowitzsch}, J., {Schreiber}, L., {Ragazzoni}, R., and {Diolaiti}, E.,
  ``{First laboratory results with the LINC-NIRVANA high layer wavefront
  sensor},'' {\em Optics Express}~{\bf 19},  16087--16095 (2011).

\bibitem{TomSpie2018}
{Herbst}, T.~M., {Arcidiacono}, C., Bergomi, M., {Bertram}, T., {Berwein}, J.,
  Bizenberger, P., Briegel, F., Fainato, J., Marafatto, L., Mathar, R.~J.,
  McGurk, R.~C., {Ragazzoni}, R., {Santhakumari}, K.~K.~R., and Viotto, V.,
  ``{Commissioning multi-conjugate adaptive optics with LINC-NIRVANA on LBT},''
  in [{\em Adaptive Optics Systems VI}{\nolinebreak\hspace{0.1em}]},  {\em This
  Conference} {\bf 1073} (2018).

\bibitem{2010SPIE.7736E..4JA}
{Arcidiacono}, C., {Bertram}, T., {Ragazzoni}, R., {Farinato}, J., {Esposito},
  S., {Riccardi}, A., {Pinna}, E., {Puglisi}, A., {Fini}, L., {Xompero}, M.,
  {Busoni}, L., {Quiros-Pacheco}, F., and {Briguglio}, R., ``{Numerical control
  matrix rotation for the LINC-NIRVANA multiconjugate adaptive optics
  system},'' in [{\em Adaptive Optics Systems II}{\nolinebreak\hspace{0.1em}]},
   {\em \procspie} {\bf 7736},  77364J (2010).

\bibitem{2014SPIE.9148E..5MB}
{Bertram}, T., {Kumar Radhakrishnan Santhakumari}, K., {Marafatto}, L.,
  {Arcidiacono}, C., {Berwein}, J., {Ragazzoni}, R., and {Herbst}, T.~M.,
  ``{Wavefront sensing in a partially illuminated, rotating pupil},'' in [{\em
  Adaptive Optics Systems IV}{\nolinebreak\hspace{0.1em}]},  {\em \procspie}
  {\bf 9148},  91485M (2014).

\bibitem{2010ApOpt..49G.174E}
{Esposito}, S., {Riccardi}, A., {Quir{\'o}s-Pacheco}, F., {Pinna}, E.,
  {Puglisi}, A., {Xompero}, M., {Briguglio}, R., {Busoni}, L., {Fini}, L.,
  {Stefanini}, P., {Brusa}, G., {Tozzi}, A., {Ranfagni}, P., {Pieralli}, F.,
  {Guerra}, J.~C., {Arcidiacono}, C., and {Salinari}, P., ``{Laboratory
  characterization and performance of the high-order adaptive optics system for
  the Large Binocular Telescope},'' {\em Applied Optics}~{\bf 49},  G174+
  (2010).

\bibitem{2012SPIE.8447E..6FM}
{Marafatto}, L., {Bergomi}, M., {Brunelli}, A., {Dima}, M., {Farinato}, J.,
  {Farisato}, G., {Lessio}, L., {Magrin}, D., {Ragazzoni}, R., {Viotto}, V.,
  {Bertram}, T., {Bizenberger}, P., {Brangier}, M., {Briegel}, F., {Conrad},
  A., {De Bonis}, F., {Herbst}, T., {Hofferbert}, R., {Kittmann}, F.,
  {K{\"u}rster}, M., {Meschke}, D., {Mohr}, L., and {Rohloff}, R.-R.,
  ``{Aligning a more than 100 degrees of freedom wavefront sensor},'' in [{\em
  Adaptive Optics Systems III}{\nolinebreak\hspace{0.1em}]},  {\em \procspie}
  {\bf 8447},  84476F (2012).

\bibitem{2004SPIE.5490.1286S}
{Soci}, R., {Ragazzoni}, R., {Herbst}, T.~M., {Farinato}, J., {Gaessler}, W.,
  {Baumeister}, H., {Rohloff}, R.-R., {Diolaiti}, E., {Xu}, W., {Andersen},
  D.~R., {Egner}, S.~E., {Arcidiacono}, C., {Lombini}, M., {Ebert}, M.,
  {Boehm}, A., {Muench}, N., and {Xompero}, M., ``{LINC-NIRVANA: mechanical
  challanges of the MCAO wavefront sensor},'' in [{\em Advancements in Adaptive
  Optics.}{\nolinebreak\hspace{0.1em}]},  {Bonaccini Calia}, D., {Ellerbroek},
  B.~L., and {Ragazzoni}, R., eds., {\em Proc. SPIE} {\bf 5490},  1286--1295
  (2004).

\bibitem{2003SPIE.4839..772B}
{Biasi}, R., {Andrighettoni}, M., {Veronese}, D., {Biliotti}, V., {Fini}, L.,
  {Riccardi}, A., {Mantegazza}, P., and {Gallieni}, D., ``{LBT adaptive
  secondary electronics},'' in [{\em Adaptive Optical System Technologies
  II}{\nolinebreak\hspace{0.1em}]},  {Wizinowich}, P.~L. and {Bonaccini}, D.,
  eds.,  {\bf 4839},  772--782 (2003).

\bibitem{2011aoel.confP..44B}
{Berwein}, J., {Bertram}, T., {Conrad}, A., {Briegel}, F., {Kittmann}, F.,
  {Zhang}, X., and {Mohr}, L., ``{End-To-End performance test of the
  LINC-NIRVANA Wavefront-Sensor system.},'' in [{\em Second International
  Conference on Adaptive Optics for Extremely Large Telescopes. Online at <A
  href=``http://ao4elt2.lesia.obspm.fr''>http://ao4elt2.lesia.obspm.fr</A>,
  id.P44}{\nolinebreak\hspace{0.1em}]},   P44 (2011).

\bibitem{2014SPIE.9148E..2YB}
{Bergomi}, M., {Viotto}, V., {Arcidiacono}, C., {Marafatto}, L., {Farinato},
  J., {Baumeister}, H., {Bertram}, T., {Berwein}, J., {Briegel}, F., {Conrad},
  A., {Kittman}, F., {Kopon}, D., {Hofferbert}, R., {Magrin}, D.,
  {Radhakrishnan Santhakumari}, K.~K., {Puglisi}, A., {Xompero}, M.,
  {Briguglio}, R., {Quiros-Pacheco}, F., {Herbst}, T.~M., and {Ragazzoni}, R.,
  ``{First light of the LINC-NIRVANA Pathfinder experiment},'' in [{\em
  Adaptive Optics Systems IV}{\nolinebreak\hspace{0.1em}]},  {\em \procspie}
  {\bf 9148},  91482Y (2014).

\bibitem{2016SPIE.9909E..2UH}
{Herbst}, T.~M., {Arcidiacono}, C., {Bertram}, T., {Bizenberger}, P.,
  {Briegel}, F., {Hofferbert}, R., {K{\"u}rster}, M., and {Ragazzoni}, R.,
  ``{MCAO with LINC-NIRVANA at LBT: preparing for first light},'' in [{\em
  Adaptive Optics Systems V}{\nolinebreak\hspace{0.1em}]},  {\em \procspie}
  {\bf 9909},  99092U (2016).

\end{thebibliography}
\bibliographystyle{spiebib} % makes bibtex use spiebib.bst

\end{document}